\documentstyle[twoside,fleqn,espcrc2,epsf]{article}

\def\tr{\mathop{\rm tr}\nolimits}
\newcommand{\dl}{\delta}
\newcommand{\Lm}{\Lambda}
\newcommand{\sg}{\sigma}
\newcommand{\Om}{\Omega}
\newcommand{\Ss}[1]{\mbox{$\cal #1$}}
\newcommand{\beq}{\begin{equation}}
\newcommand{\bea}{\begin{eqnarray}}
\newcommand{\eeq}{\end{equation}}
\newcommand{\eea}{\end{eqnarray}}
\newcommand{\hf}{\frac{1}{2}}
\newcommand{\half}{\sfrac{1}{2}}

\newcommand{\drs}{{three-sphere}}
\newcommand{\ie}{{i.e.\ }}

\newcommand{\Next}{\nonumber \\ }
\newcommand{\pr}{\partial}
\newcommand{\Sdr}{{S$^3$}}
\newcommand{\SUtw}{{SU(2)}}
\newcommand{\EUC}{{\mbox{\scriptsize E}}}
\newcommand{\Skl}{{\mbox{\scriptsize S}}}
\newcommand{\un}{\underline}
\newcommand{\Order}[1]{\Ss{O}\left(#1\right)}
\newcommand{\sfrac}[2]{\mbox{\large $\frac{#1}{#2}$}}

\newcommand{\AmS}{{\protect\the\textfont2
  A\kern-.1667em\lower.5ex\hbox{M}\kern-.125emS}}

\title{\vskip -1cm   \hfill{\small INLO-PUB-13/96} \vskip 5mm
Glueballs on \Sdr}

\author{Bas van den Heuvel\address{
Instituut-Lorentz for Theoretical Physics, University of Leiden, \\
  PO Box 9506, NL-2300 RA Leiden, The Netherlands}
}

\begin{document}
\begin{abstract}
For \SUtw\ gauge theory on the \drs\ we study the dynamics 
of the low-energy modes.  
By explicitely integrating out the high-energy modes, 
the one-loop correction to the hamiltonian for this problem is obtained.
After imposing the $\theta$ dependence
through boundary conditions in configuration space, we
obtain the glueball spectrum of the effective theory with a
variational method.
\end{abstract}

\maketitle

\section{INTRODUCTION}
It is our goal to study the influence of the structure
of the configuration space on the \SUtw\ glueball spectrum.
The method we use consists of putting the theory in a
finite spatial volume and reducing the full dynamics
to an effective theory of a finite number of degrees of 
freedom~\cite{lus2,baa2}.
Asymptotic freedom implies that
a small volume corresponds to a small coupling constant.
Hence changing the volume from small to intermediate values
allows us to monitor the onset of non-perturbative phenomena.
\par
The Yang-Mills configuration space is the space of gauge orbits
$\cal{A}/\cal{G}$ ($\cal{A}$ is the collection of 
gauge fields or connections, $\cal{G}$ the group of local gauge transformations).
We know from Singer~\cite{sin} that the topology of this configuration space
is highly non-trivial when \Ss{G} is non-abelian.  
At increasing coupling, the wave functional will start to 
spread out and will become
sensitive to the non-trivial topology, like
non-contractable closed loops, of the configuration space.
\par
This spreading out will occur first in those directions in
configuration space, where the potential rises the slowest,
like in the direction of the low-energy modes of the gauge field,
or in the direction of the sphalerons associated to tunnelling
from one Gribov copy of the vacuum to another.
As long as the global features of the configuration space only affects
this finite number of modes, one 
can capture these non-perturbative phenomena in an effective model of the
low-energy modes.
One uses the Born-Oppenheimer approximation to account for 
the influence of all the other (high-energy)
modes, that is, they are assumed to behave perturbatively and
are integrated out from the path integral.
\par
We are interested in the influence of the multiple vacuum structure of
the theory on the glueball spectrum; in particular we would like 
to see the dependence of the energies on the $\theta$ angle.
The $\theta$ angle
shows up when one implements gauge invariance under large 
gauge transformations. If $n[g]$ is the winding number of
the gauge transformation $g$, gauge invariance in the hamiltonian
formulation is implemented by
\beq
  \Psi[^g \! A] = e^{ i n[g] \theta}~\Psi[A].
\eeq
We impose
gauge invariance by restricting the theory to a 
so-called fundamental domain~\cite{sem,zwan}. Here the
$\theta$ angle shows up in the boundary conditions that have 
to be imposed at the boundary of the fundamental domain.
\par
To compare results with lattice calculations it would be most
natural to take the finite (spatial) volume to be a 3-torus T$^3$.
On the torus, however, the instantons, which are
the gauge field configurations that describe tunnelling between 
different vacua, are only known numerically~\cite{gar}.
To circumvent this problem
we take our space to be the three-sphere \Sdr~\cite{cut1,baa16},
where the instantons are exactly known. The sphaleron degrees
of freedom for this geometry happen to be embedded in the
space of the low-energy modes.

\section{THE EFFECTIVE THEORY}
\begin{figure}[t]
\epsfxsize=\columnwidth
\epsffile{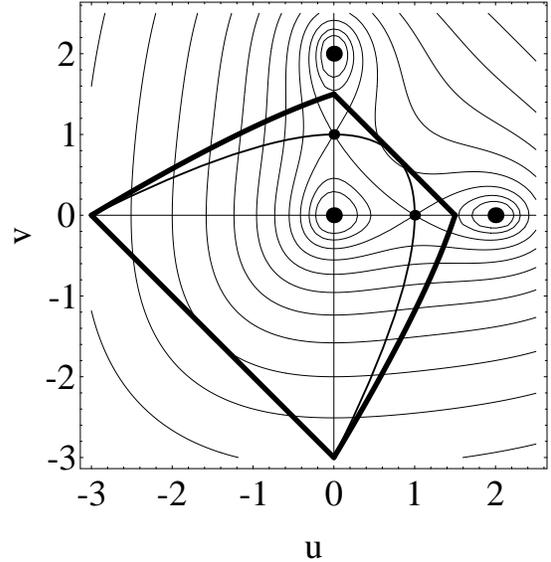}
\caption{
The $(u,v)$ plane: $c^a_i = - u \dl^a_i,~d^a_i = - v \dl^a_i$.
Location of the classical vacua, sphalerons, lines of equal potential.
The boundary of the fundamental domain lies between the Gribov horizon 
(fat sections) and the
lower bound $\tilde{\Lm}$ (drawn parabola): $\tilde{\Lm} \subset
\Lm \subset \Om$, with $\Om$ the Gribov region.}
\label{uvplane}
\end{figure}

To isolate the low-energy modes, we examine the potential energy
and the quadratic fluctuation operator \Ss{M} defined by
\bea
 V(A) &=& - \frac{1}{2 \pi^2} \int_{\Skl^3} \half \tr(F_{ij}^2) \Next
 &=& - \frac{1}{2 \pi^2} \int_{\Skl^3} \tr(A_i \Ss{M}_{ij} A_j) 
   + \Order{A^3}. \label{fluctdef}
\eea
The space of low-energy modes is the lowest eigenspace of \Ss{M}.
It is 18 dimensional and can be parametrized by~\cite{baa16}
\beq
    A_\mu(c,d) = 
  \left(c^a_i  e^i_\mu+d^a_j\bar{e}^j_\mu \right) 
  \frac{\sg_a}{2},
\label{Bcddef}
\eeq
with $\sg_a =  i \tau_a$, $\tau_a$ the Pauli matrices, and
$e^i_\mu$ and $\bar{e}^i_\mu$ certain framings on \Sdr.
We now have to restrict ourselves to the cross section of this space 
with the fundamental domain $\Lm$ and then solve for the lowest eigenfunctions 
of the hamiltonian. To make this problem well-defined we have to specify 
boundary conditions. To keep things transparent we now 
focus on a subspace of the 
18-dimensional space.
This two-dimensional space of field configurations (see fig.~\ref{uvplane})
contains three copies of the vacuum (large dots),
which are related by large gauge transformations.
This space is important because
of all the tunnelling paths connecting the vacua, it contains
those paths that have the lowest energy barrier. 
These barrier configurations are saddle points of $V$ and are called
sphalerons (small dots); they are also gauge
copies of each other.
At the regime
we start to see the non-perturbative effects,
that is, where energies become of the order of the
sphaleron energy, the only relevant
boundary conditions~\cite{heu1} are those at the sphalerons:
\beq
  \Psi(A(1,0)) = e^{i \theta} \Psi(A(0,1)).
\eeq
The lowest order (or truncated) hamiltonian becomes:
\beq
  -\frac{f}{2} \left( 
  \frac{\pr^2}{\pr c^a_i \pr c^a_i} + \frac{\pr^2}{\pr d^a_i \pr d^a_i} 
   \right) + \frac{1}{f} \frac{\Ss{V}(c,d)}{2 \pi^2},
  \label{hamil}
\eeq
with $f = \frac{g(R)^2}{2 \pi^2}$ the renormalized coupling constant.
\par
The potential $\Ss{V}(c,d)$ for the truncated hamiltonian is obtained
directly from eq.~(\ref{fluctdef}).
The one-loop correction to the hamiltonian is obtained
by explicitly integrating out the high-energy modes~\cite{heu2}, 
that is, the  non-$(c,d)$ modes in the path integral.
Using background gauge fixing, $D(B(c,d))_\mu Q_\mu = 0$,
we find for the effective euclidean action:
\beq
  -S^{\mbox{eff}}_{\EUC}[B] = 
  -S^{\mbox{cl}}_{\EUC}[B]
  -S^{\mbox{eff}(1)}_{\EUC}[B],
\eeq
with the one-loop contribution $-S^{\mbox{eff}(1)}_{\EUC}[B]$ given by
\bea
&& \ln \int D\psi~D\bar{\psi} DQ_\mu \exp\left[ 
  \int d \tau \int_{\Skl^3} d \vec{x} \times \right. \Next 
&& \quad
\left. \tr\left( \bar{\psi} (-D_\mu^2(B)) \psi
 + Q_\mu W_{\mu \nu}(B) Q_\nu \right)
 \vphantom{\int} \right] = \Next
&& \ln \left[ {\det}'(-D_\mu^2(B)) / {{\det}'}^{\half}(W_{\mu \nu}(B))\right].
\eea
Here we introduced ghost fields $\psi$ and $\bar{\psi}$, and
$W(B)$ denotes the inverse field propagator in the background field $B$.

\section{THE RAYLEIGH-RITZ ANALYSIS}

We approximate the excitation energies of the effective model,
which are the masses of the various glueballs, with a
variational method. We introduce the radial coordinates
\beq
  r_c = \left[ c^a_i c^a_i \right]^\hf, \quad 
  r_d = \left[ d^a_i d^a_i \right]^\hf.
\eeq
We rewrite the hamiltonian of eq.~(\ref{hamil}) in radial and
angular coordinates. Remembering the remarks above on the relevance
of the boundary conditions, we can implement them through
(see~\cite{heu1} for details)
\beq
\left\{
\begin{array}{ll}
  \psi(\mbox{Sph},0) &= e^{i \theta} \psi(0,\mbox{Sph}) \\
  \frac{\pr (r_c^{-\frac{5}{2}}\psi)}{\pr r_c}(\mbox{Sph},0) &=  
    - e^{i \theta} \frac{\pr (r_d^{-\frac{5}{2}}\psi)}{\pr r_d}(0,\mbox{Sph})
\end{array}
\right.
\eeq
The trial wave functions we use for the Rayleigh-Ritz method~\cite{ree}
are essentially the eigenfunctions of the kinetic part of the 
hamiltonian. The rotational symmetry is used to classify the
various glueball states.

\section{RESULTS}

\begin{figure}[t]  \centering
  \epsfxsize=\columnwidth
  \leavevmode
  \epsfbox[72 222 540 570]{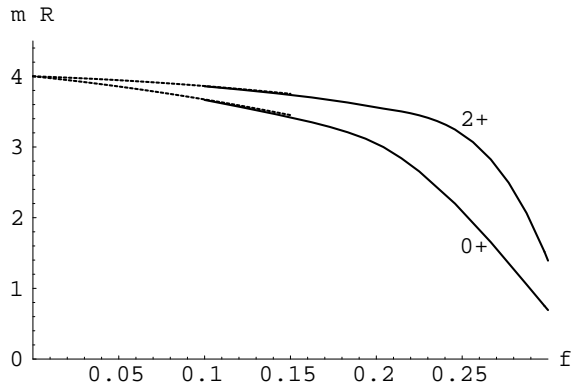}\\ 
  \caption{One-loop results. 
  Glueball masses for $\theta = 0$ as a function of 
  the coupling constant. The drawn curves are the masses of the 
  first scalar ($0^+$) and tensor ($2^+$) glueball. The dotted
  lines denote the perturbative results.}
  \label{massasone}
\end{figure}
Results for the mass of the first scalar and tensor glueball
are shown in fig.~\ref{massasone}. For small coupling 
these masses can be calculated perturbatively, but at
$f \approx 0.2$ we see the onset of the influence of the boundary
conditions, \ie of the instantons. Beyond $f \approx 0.3$, our
model will no longer be valid, because the wave function
will no longer be sensitive to just the boundary conditions at
the sphaleron, as can be checked explicitly using plots
of the wave function. More results, including the dependence
on the $\theta$ angle, can be found in \cite{heu1,heu2}
\par
It is important to emphasize one should not expect our results for
the spectrum to be accurate for large volumes, but it has been
the main aim of this study
to demonstrate that instanton effects on the low-lying spectrum
are large, but calculable as long as energies remain close to the 
sphaleron energy, where nevertheless semiclassical techniques will
completely fail.

\end{document}